\documentclass[3p,twocolumn,pra,aps,showpacs,superscriptaddress,article]{elsarticle}
\usepackage[english]{babel}
\usepackage[utf8]{inputenc}
\usepackage{amsmath,amssymb,amsthm}
\usepackage{xcolor}
\usepackage{float}
\usepackage{textcomp}
\usepackage{amssymb}
\usepackage{graphicx}
\usepackage{esint}
\usepackage{lineno,hyperref}
\usepackage[toc,page]{appendix}
\def\<{\left\langle}
\def\>{\right\rangle}
\def\({\left(}
\def\){\right)}
\def\dis{\textrm{dis}}
 \journal{Physica A}

\begin{document}

\begin{frontmatter}

\title{A Schelling model with a variable treshold in a closed city segregation model. Analysis of the universality classes.}

\author{Diego Ortega}
\ead{dortega144@alumno.uned.es}
\address{Dto. Física Fundamental, Universidad Nacional de Educacion a Distancia (UNED), Spain}

\author{Javier Rodríguez-Laguna}
\address{Dto. Física Fundamental, Universidad Nacional de Educacion a Distancia (UNED), Spain}

\author{Elka Korutcheva}
\address{Dto. Física Fundamental, Universidad Nacional de Educación
  a Distancia (UNED), Spain}
\address{G. Nadjakov Institute of Solid State Physics, Bulgarian Academy of Sciences, 1784 Sofia, Bulgaria.}

\begin{abstract}
Residential segregation is analyzed via the Schelling model, in which
two types of agents attempt to optimize their situation according to
certain preferences and tolerance levels. Several variants of this
work are focused on urban or social aspects. Whereas these models
consider fixed values for wealth or tolerance, here we consider how
sudden changes in the tolerance level affect the urban structure in
the closed city model. In this framework, when tolerance decreases
continuously, the change rate is a key parameter for the final state
reached by the system. On the other hand, sudden drops in tolerance
tend to group agents into clusters whose boundary can be characterized
using tools from kinetic roughening. This frontier can be categorized 
into the Edward-Wilkinson (EW) universality class.
Likewise,  the understanding of these processes and how
society adapts to tolerance variations are of the utmost importance in a world where
migratory movements and pro-segregational attitudes are commonplace.

\end{abstract}


\end{frontmatter}


\section{Introduction}

People with similar features (culture, income, etc.) tend to group
together in the same neighborhood, giving rise to segregation on a
social scale. More than 40 years ago, Schelling put forward a seminal
model that describes this reality, linking individual preferences to
the macroscopic behaviour of the system \cite{key-1p}. Two different
social groups who usually represent people with different culture or income are assigned two colors: {\em red} and {\em blue}. These groups are
distributed over a square lattice with some vacancies on it. Agents
are characterized by a {\em tolerance} $T$: the fraction of different
agents in their neighborhood that he or she can tolerate. The model
proceeds through the following dynamical rules: a random agent $i$ is
selected and his/her fraction of diverse neighbors is evaluated. If
this fraction value is lower or equal than $T$, the agent remains at
his/her location. Otherwise, he/she relocates to the nearest vacancy
that meets his/her demands. For intermediate values of $T$ we observe
segregation, and clusters are formed with different types of agents.

This model has attracted a great deal of attention, due to its
simplicity and insight, giving rise to a wealth of variants. System
behaviour when one kind of agents are tolerant and the vacancies are
differently priced was characterized in \cite{key-2p}. Differences
between constrained models, where only {\em unhappy} individuals are
allowed to move, and unconstrained ones, in which all agents can
relocate to vacancies as long as they keep or increase their {\em
happiness}, were also studied \cite{key-3p}. In \cite{key-4p},
attempted relocations succeed with a probability which is modulated by
a power-law linking their current happiness and the attractiveness of
the offered place. The effect of the city shape, size and form is
investigated in \cite{key-5p}, finding that the properties of the
system in equilibrium are weakly affected by these parameters. The
authors of \cite{key-6p} proposed a thermodynamic approach to
segregation based on their cluster geometry, and considered quantities
analogous to the specific heat and susceptibility, along with a
connection with spin-1 models. Recently, the use of different tolerance
levels for the agents was proposed in \cite{key-7p}, in a system with
no vacancies, where agents could only exchange locations with agents
of a different type. On the other hand, in \cite{key-8p} each cell of
the system is considered a building containing many agents, and
segregation was considered both at a microscopic and a macroscopic
level, giving rise to a complex phase diagram. Some of the mentioned
works take into account the importance of the initial conditions
\cite{key-2p,key-7p}, and some others also consider migratory
movements \cite{key-7p,key-9p}.

In this paper we consider a closed city model, in which no
agents can enter or leave, and how it adapts to decays in the
tolerance level. We consider two types of decline: a sudden drop,
which may be linked to a specific violent event, and also a
continuous decay, which could be associated to a continuous flow of
biased news. In this latter case, the speed of the process is a key
parameter in order to describe the final system. In the case of a
sudden drop, two clusters may emerge, separated by an interface
where the vacancies concentrate.  This boundary, which belongs to the Edwards-Wikinson universality class, is strongly linked to the random deposition model with surface relaxation (RDSR) \citep{family_scaling_1986}. Our general framework is established in similarity with the Blume-Emery-Griffiths (BEG) \cite{key-11p}.

The paper is organized as follows. In Section \ref{sec:model} we
define our model and discuss the dynamics and the evolution process. In Section \ref{sec:results} we describe our results, linking them to well-known social mechanisms. Our main conclusions and proposals for further work are discussed in section \ref{sec:conclusion}. The connection between our model and the BEG model is established in \ref{app}.


\section{Model}
\label{sec:model}

Let us consider two kinds of agents to be living in an $N\times N$ square lattice with open boundaries (non-periodic) and a fixed vacancy density $\rho$, in similarity to \cite{key-6p}. Agents will not be allowed to enter or leave the system. The key parameter of the system is the tolerance $T$,  the fraction of different
agents in their neighborhood that an agent can tolerate while staying happy.  It is mathematically defined as:

\begin{equation}
  T=N_d(i)/[N_s(i)+N_d(i)],
  \label{eq:1}
\end{equation}

being $N_s$ and $N_d$, respectively, the number of neighboring agents of the same (s) and different (d) type. This value is the same for all agents in the model, irrespective of their kind.

In order to make an explicit connection between our physical model and social realities, we define a measure of the level of unhappiness of an agent. The lack of happiness of agent $i$ is measured by the {\em dissatisfaction} index $I_\dis(i)$:

\begin{equation}
  I_\dis(i)=N_d(i)-T[N_s(i)+N_d(i)],
  \label{eq:2}
\end{equation}
where the neighborhood of any agent is comprised by a maximum of his/her eight closest neighbors, given that we are considering a Moore neighborhood. This number is reduced for agents on the system edges, since the lattice is defined with open boundary conditions which take into account the finite size of cities. It should be pointed out that smaller values of $I_\dis(i)$ correspond to happier agents.

The system dynamics is as follows: we start from a random initial configuration with the same number of red and blue agents. At each iteration, a random occupied site $i$ and a random vacancy $j$ are selected. The value of $I_\dis(j)$ is calculated using Eq. \eqref{eq:2}. The proposed relocation is accepted if $I_\dis(j)\leq 0$. The previous condition implies $happiness$ for the agent in the new location. However, we must note that a relocation of an agent to another place where his/her happiness level is inferior is possible, if the destination environment verifies $T\geq N_{d}(j)/\left(N_{s}(j)+N_{d}(j)\right)$. This fact induces further relocations in the system causing that a final static equilibrium state is never reached.

We have considered two different schedules for the decrease in tolerance: a continuous variation of the tolerance threshold and a sudden drop. The former can be associated with biased news, and the tolerance value is described by a decreasing function since the beginning of the simulation. The latter can be related to some extreme form of violence. In this case the value of $T$ is decreased abruptly, at certain time step, after the system has reached equilibrium. In this simulation two levels of tolerance are considered before the sudden drop: an intermediate level, $T=1/2$, and an upper one with $T=7/8$. Our purpose is to study how the initial tolerance level affects the system final state.

The parameters of the model  are the system size $N$,
the tolerance level $T$ and the vacancy density $\rho$. Typical values
for $\rho$ in our simulations are under $0.1$, given that cities are 
densely populated.


\section{Results and discussion}
\label{sec:results}

As we have discussed in Sec \ref{sec:model}, two equal populations of
agents inhabit an $N\times N$ square lattice. We will use $N=50$ and a
fixed vacancy ratio $\rho=6\%$, unless otherwise specified. Agents can
not enter or leave the system, i.e. no external changes are
allowed. However, an agent $i$ is able to move into an empty cell $j$,
randomly offered, if $I_\dis(j)\leq 0$. Note that the relocation
process may increase the dissatisfaction index of some of the old or
new neighbors of the chosen agent.

A phase diagram for different values of $T$ and $\rho$ was presented in
\cite{key-6p}, which we will take as our starting point. Once the
density $\rho$ is fixed, $T$ remains as the only control parameter of
the system. Beginning with a random initial configuration and depending on $T$, the system can be found in three different states, which we will call {\em frozen}, {\em  segregated} and {\em mixed}. Each of them is characterized by  some stationary morphologies and the acceptance rate for relocations.

\begin{enumerate}
\item Low $T$, or {\em frozen}. Few changes are accepted and the
  system remains close to the random initial configuration: an aleatory
  mixture of red agents, blue agents and vacancies.
\item Medium $T$ or {\em segregated}. Two big clusters are created and
  the accepted change rate is close to 50\% in equilibrium.
\item High $T$ or {\em mixed}. Almost all changes are accepted, so no
  clusters are formed and the configuration remains close to random.
\end{enumerate}

\subsection{Continuous evolution of T}

As it was discussed previously, we will focus on how the system adapts
to changes in tolerance. So, first, we characterize the system
behavior when the tolerance value $T$ decreases according to the
following law

\begin{equation}
  T(t)=1-\tanh\(\frac{t}{t_0}\),
  \label{eq:10}
\end{equation}
where $t$ is the time measured in Monte Carlo steps and $t_0$ is a
factor controlling the overall speed of the process. This kind
of gradual tolerance decrease can be associated with a sustained
process of biased news or a gradual rising of pro-segregational
parties. Eq. \eqref{eq:10} describes the evolution of the
tolerance in a city that changes from being extremely tolerant
($T\sim1$) to totally intolerant ($T\sim 0$). Futhermore, it is also
possible to consider the tolerance $T$ as an analogue of the system
temperature \cite{key-6p}, so the process can also be understood as a
cooling process.

We will put special emphasis on the sizes of the different clusters,
measured through the segregation coefficient \cite{key-13p}, $s.c.$,
given by

\begin{equation}
  s.c.=\frac{2}{N^4\rho)^2}\sum_{\left\{ c\right\} }n_c^2,
  \label{eq:9}
\end{equation}
where $c$ indexes all the clusters in the system and $n_c$ is the
number of agents in each cluster. This coefficient ranges from values
close to 0, where clusterization has not taken place, to 1, where only
two clusters remain and $n_c=N^2(1-\rho)/2$.

\begin{figure}[H]
\begin{center}
\includegraphics[width=8cm]{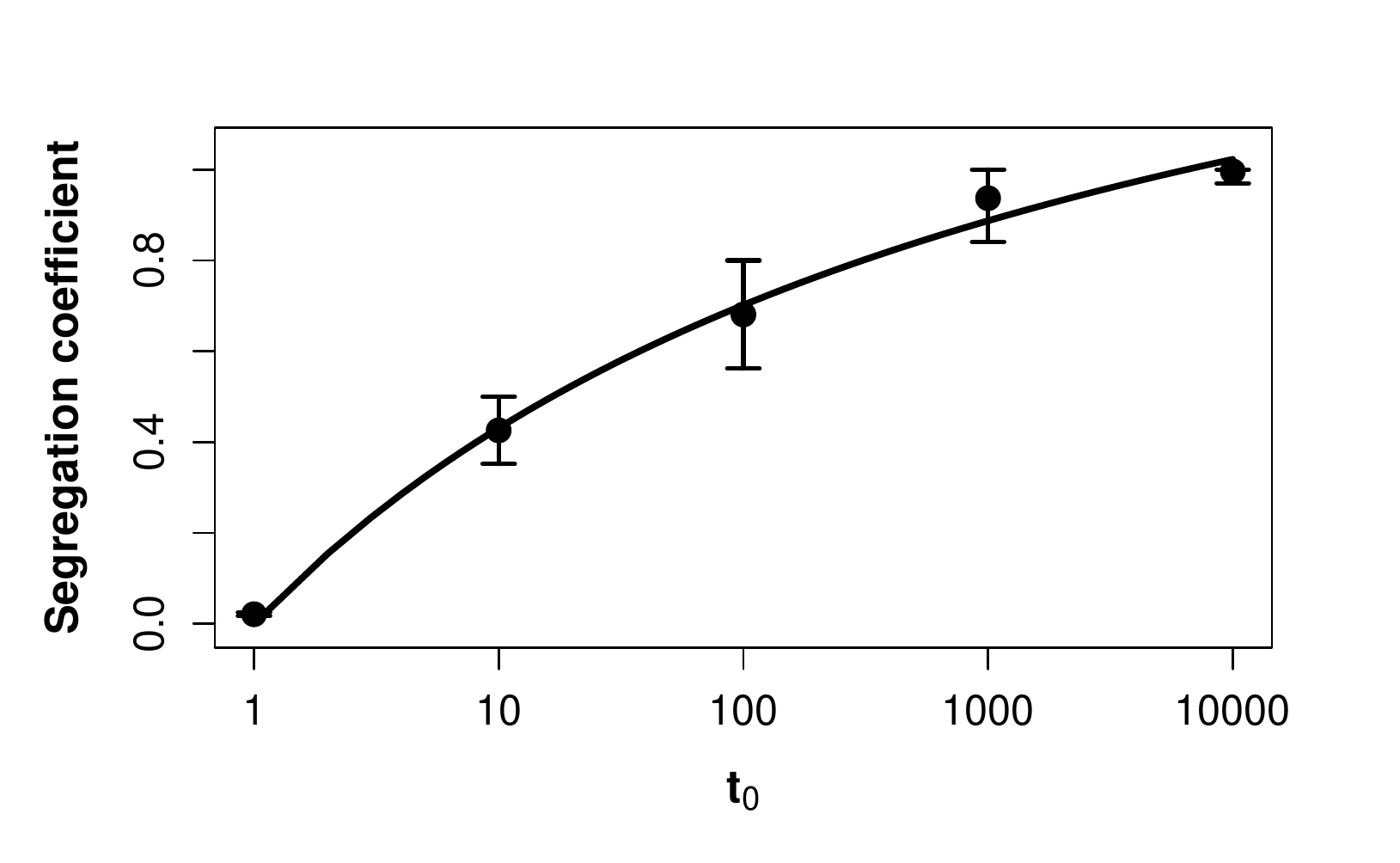}
\end{center}
\caption{Values of final segregation coefficient for different $t_0$ values. Each point correspond to the average over 50 runs, while error bars equal standard deviations. Final considered times are up to $30t_0$.}
\label{fig1}
\end{figure}

The value of the final segregation coefficient for different values of
$t_0$ ranging from $1$ to $10^4$ is shown in Fig. \ref{fig1}. Starting out from an intial random configuration, as $t_0$ increases, the system spends more time on intermediate values of $T$, rising the clusterization effect. As a consequence, the segregation coefficient becomes larger when $t_{0}$ increases. For $t\gg t_0$, $T$ becomes very low, the lattice freezes and the structure created in the previous stage remains. Time is measured in Monte Carlo (MC) steps, corresponding to $N\times N$ iterations of the system dynamics.

\begin{figure}
\centering
\begin{tabular}{cc}
\includegraphics[width=3.5cm]{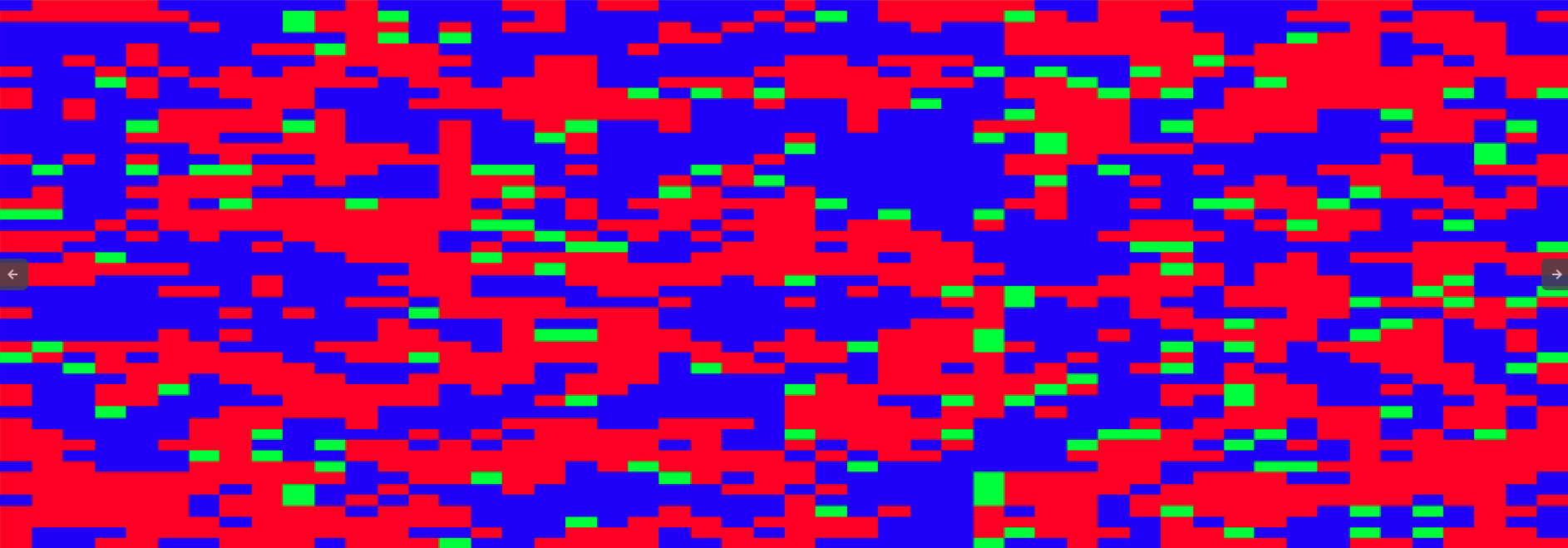} &
\includegraphics[width=3.5cm]{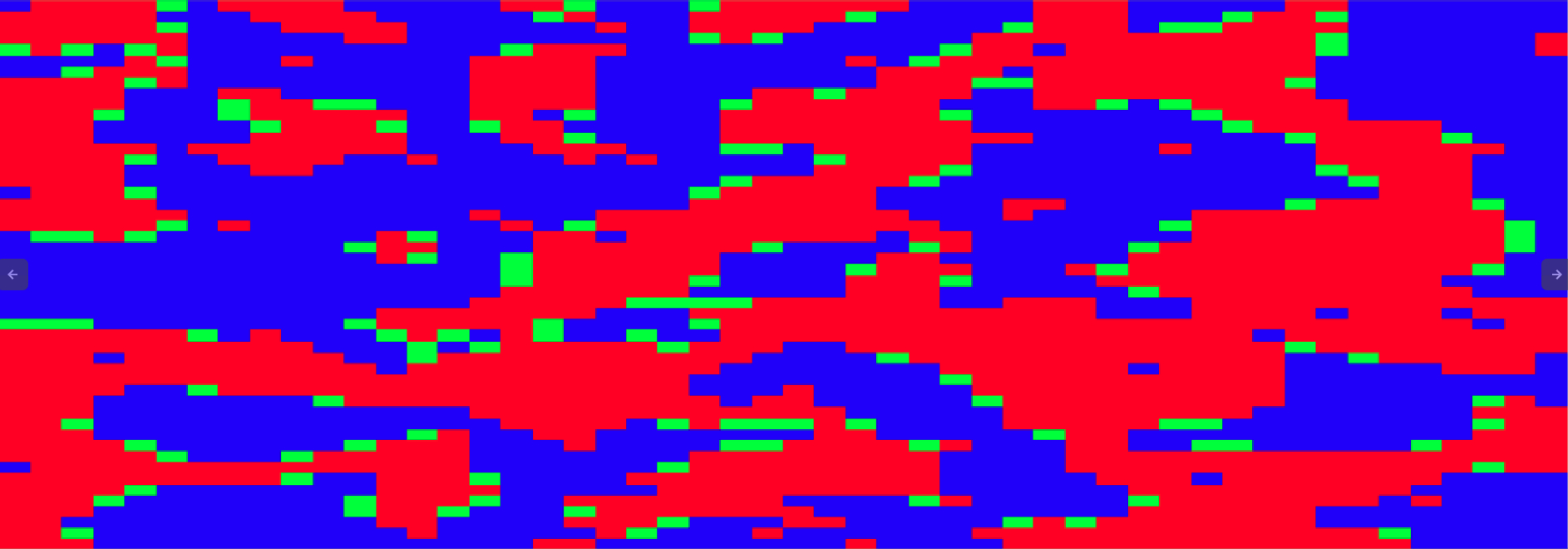} \\
(a) & (b)\\
\includegraphics[width=3.5cm]{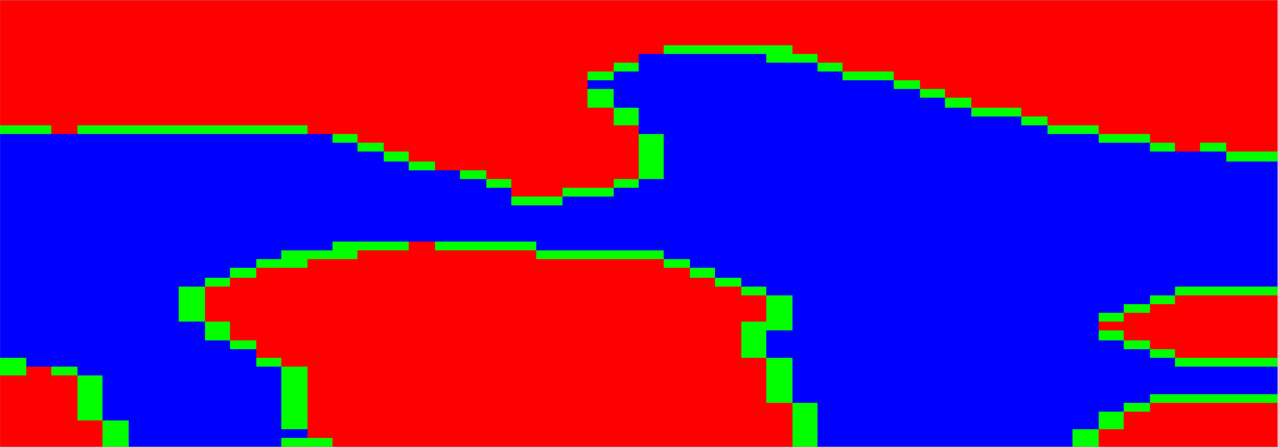} &
\includegraphics[width=3.5cm]{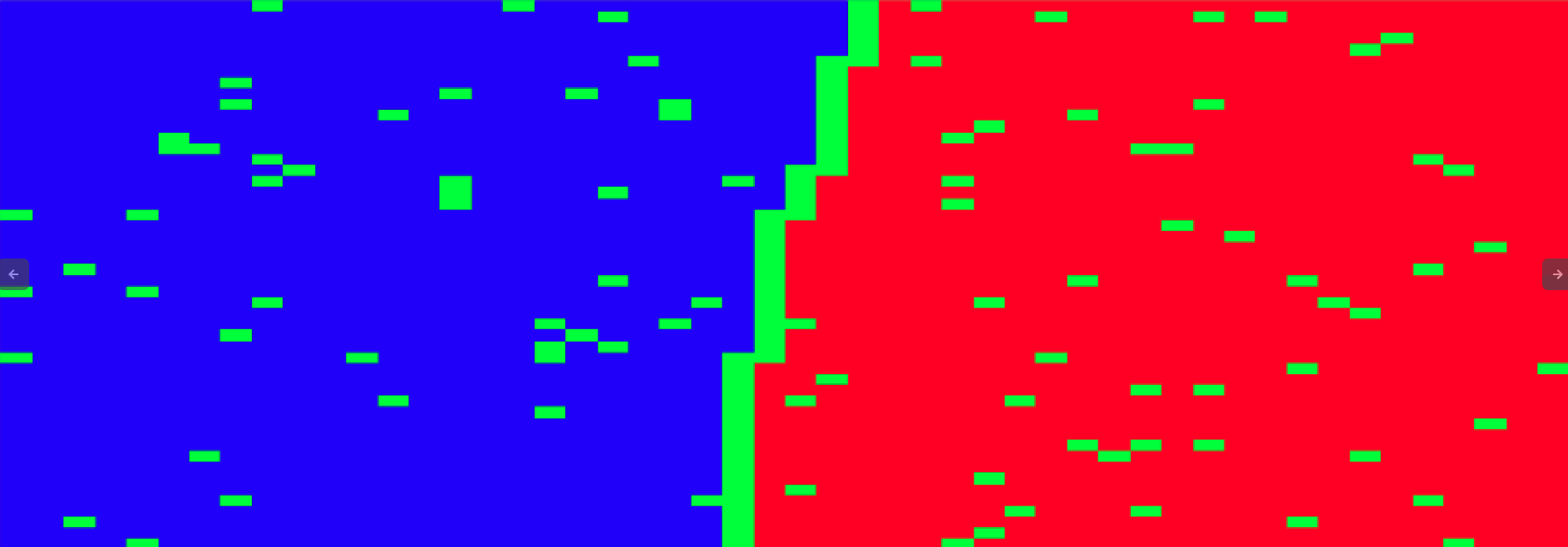} \\
(c) & (d) \\
\end{tabular}
\caption{Snaphshots of the system final state for $t_{0}$ values:
  1 (a), 10 (b), 100 (c) and 1000 (d). The snapshot for $t_{0}=10^4$ is similar to (d), so it has not been shown.}
\label{fig2}
\end{figure}

As we can see from snapshots of the final system state in Fig.\ref{fig2}, segregation varies greatly between $t_{0}=100$ and $t_{0}=1000$. Besides, the value of $s.c.$ increases from $0.42$ to $0.68$, respectively (Fig. \ref{fig1}). Therefore, we may claim that the transition towards a segregated state takes place at some point within this range.

From a social perspective, a slow decrease in tolerance
forces society to establish big clusters of differentiated groups,
minimizing their contact zones. Likewise, when the process becomes
faster the reorganization process becomes less efficient, a large
number of small clusters are formed and the total size of the
boundary between both groups rises, thus increasing the
likelihood of conflicts.

\subsection{Sudden change in T}

Starting with an initial random configuration, we fix the tolerance value and let the system evolve until a new equilibrium is attained. After a long time has passed, tipically $t=21000$ MC steps, we impose a sudden drop in the tolerance towards a value of $T=1/4$, within a single MC step. This kind of decrease in tolerance could be associated with localized violent events with a great impact on public opinion. Two types of societies are considered, depending on the initial value of the tolerance denoted as $T_i$: one is highly tolerant, with $T_i=7/8$, and the other one is
segregated, with $T_i=1/2$. We have observed a phenomenon that also
takes place under some conditions for the continuous variation: after
the change in $T$, the vacancies of the system are grouped at the
interfaces between red and blue agents, see Fig. \ref{FIG4} (a) and
(b).

Let us consider a vacancy at a flat segment of the border between red
and blue agents. We can calculate the critical tolerance value for its
location as $T^*=4/(4+4)=1/2$. Before the drop in tolerance, $T=1/2$,
so agents can accept to move into these vacancies. After the drop,
when $T=1/4$, no relocation can take place anymore towards this place,
because $T<T^*$, so the vacancy must remain empty. After a short time,
most vacancies inside clusters have been transferred into the
interface and it becomes flat. From a social perspective, nobody wants
to be placed between two intolerant social groups: red agents find
this place too close to the blue ones and vice versa. 

\begin{figure}[H]
\begin{center}
\includegraphics[width=2cm]{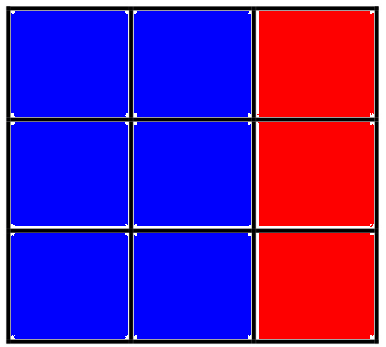}
\end{center}
\caption{Diagram of a flat segment between two clusters, centered in a blue agent }
\label{FIG3}
\end{figure}

\begin{figure*}
\begin{center}
\begin{tabular}{c}
\includegraphics[width=3cm]{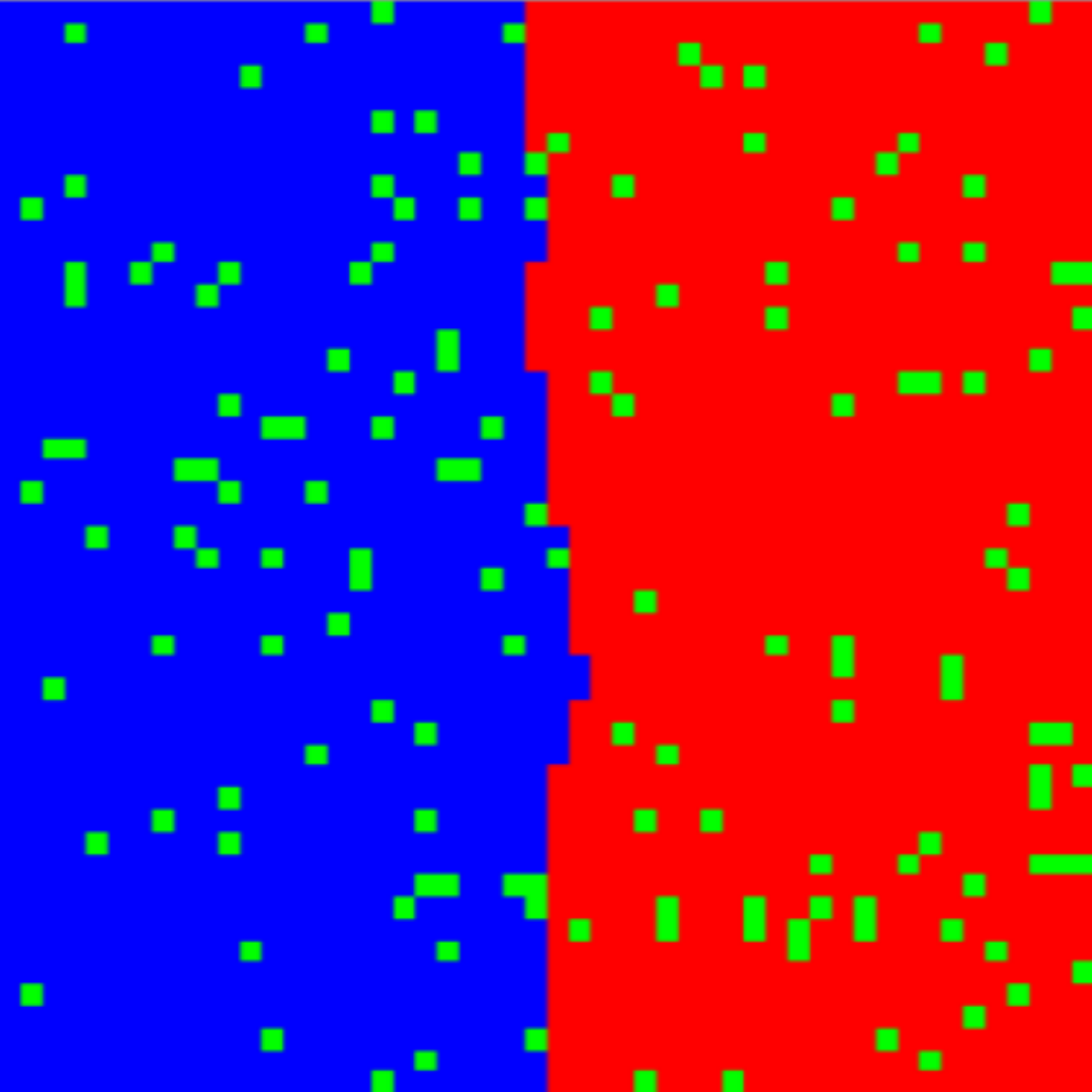}\\
(a)\\
\includegraphics[width=3cm]{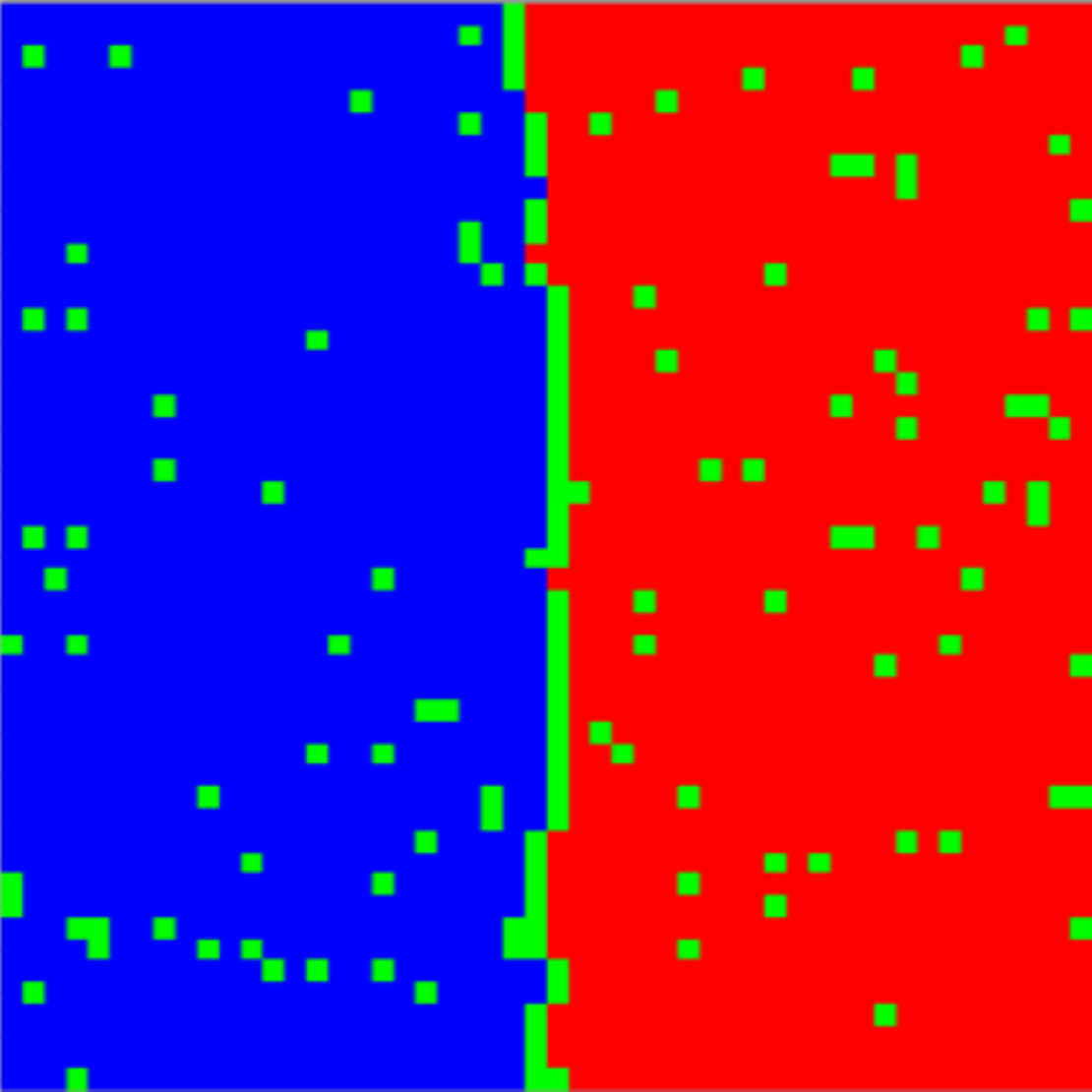}\\
(b)\\
\end{tabular}
\begin{tabular}{c}
\includegraphics[width=8cm]{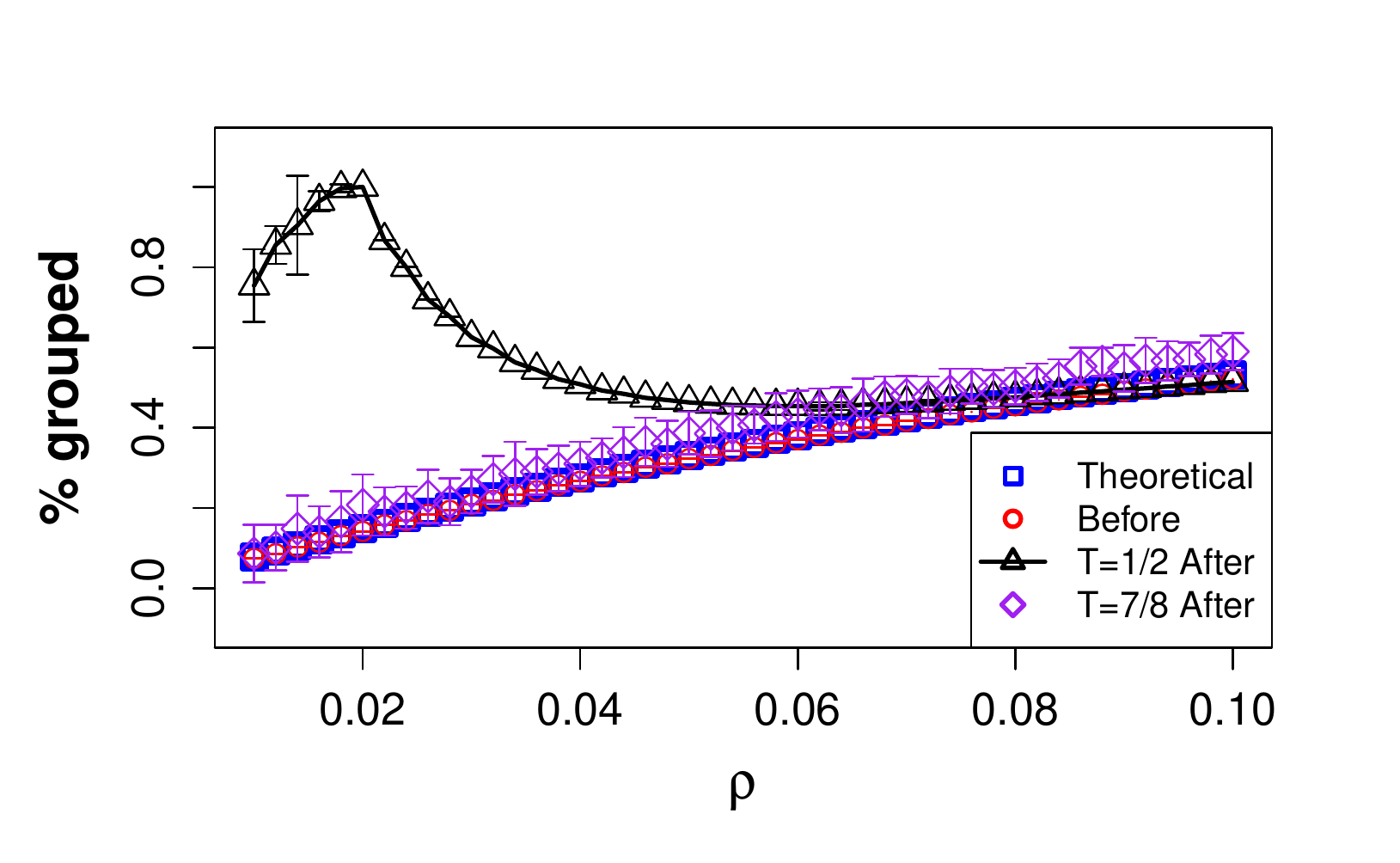}\\
(c)\\
\end{tabular}
\end{center}
\caption{Long term configurations for (a) $T=0.5$ and (b) $T=0.25$ for
  $N=50$ and $\rho=6\%$. Green squares represent vacancies. (c) Grouped vacancy ratio as a function of
  $\rho$. 50 runs and 20000 iterations are considered for each run.}
\label{FIG4}
\end{figure*}

The treshold for the creation of the vacancy interface is $T=3/8$, meaning that below this value the vacancy border will appear. This $T=3/8$ can be obtained by taking into account geometrical arguments. Let us consider a blue agent surrounded by 8 neighbors: 5 blue ones and 3 red ones. This is the situation for a flat interface as in Fig. \ref{FIG3}. When $T<3/8$, see Eq. \ref{eq:2}, $I_{dis}>0$ and the agent leaves its place. No other agents find the vacancy in this spot suitable, so it remains empty.

In order to characterize the interface we define the {\em grouped
  vacancy ratio}, which is defined as the proportion of vacancies that
have either one or two more vacancies in their neighborhood.

We can provide a theoretical estimate for this magnitude as a function
of the population density ratio. At the beginning of the simulation the vacancies are randomly distributed. Then, as the system evolves, there are not preferential aggregation for vacancies over one kind of agents, so their distribution remains random.  It is only after the drop in tolerance that they are grouped into the border between clusters. Therefore, before the cooling, we can assume a binomial distribution for the presence of one or two vacancies among the eight cells that comprise the Moore neighborhood.

\begin{equation}
\left(\begin{array}{c}
8\\ 1
\end{array}\right)\rho(1-\rho)^7+\left(\begin{array}{c}
8\\ 2
\end{array}\right)\rho^2(1-\rho)^6.
\label{eq:11}
\end{equation}

The grouped ratio is evaluated in four different situations in
Fig. \ref{FIG4} (c). Measurements before the drop in $T$ are shown as
red circumferences, which can be compared with the theoretical estimates,
shown as hollow blue squares. Measures after the drop are shown for $T=1/2$
using black triangles and for $T=7/8$ using purple diamonds. After the
drop in $T$, the grouped ratio presents major differences between the
segregated society (black traingles, $T=1/2$) and the mixed society
(purple diamonds, $T=7/8$). As we can see in Fig. \ref{FIG4} (b), a
straight line dividing the system into two big clusters requires at
least $N$ vacancies, i.e., the system length. The corresponding value
of $\rho$ is, therefore, $\rho=1/N$. In our case, $N=50$ and the
maximum grouped ratio value reached corresponds to $\rho=0.02$, as we
can see in Fig. \ref{FIG4} (c). If there are less vacancies, they will
be randomly placed along the boundary between the clusters. However,
when $\rho>1/N$, some vacancies may diffuse into the bulk of the
clusters, while some other vacancies may allow the boundary to become
rough. For $\rho\gg 1/N$ most vacancies are located in the bulk, and
the grouped ratio approach the predictions following the binomial
distribution.

For $T_i=7/8$ (purple diamonds in Fig. \ref{FIG4} (c)) these
arrangement effects on the boundary become negligible because
clusters are not actually formed. Thus, the vacancies can be reordered
during the process, but the grouped value only rises slightly along
the procedure.

\subsection{Characterization of the boundary}
\begin{figure}[H]
\begin{center}
\includegraphics[width=8cm]{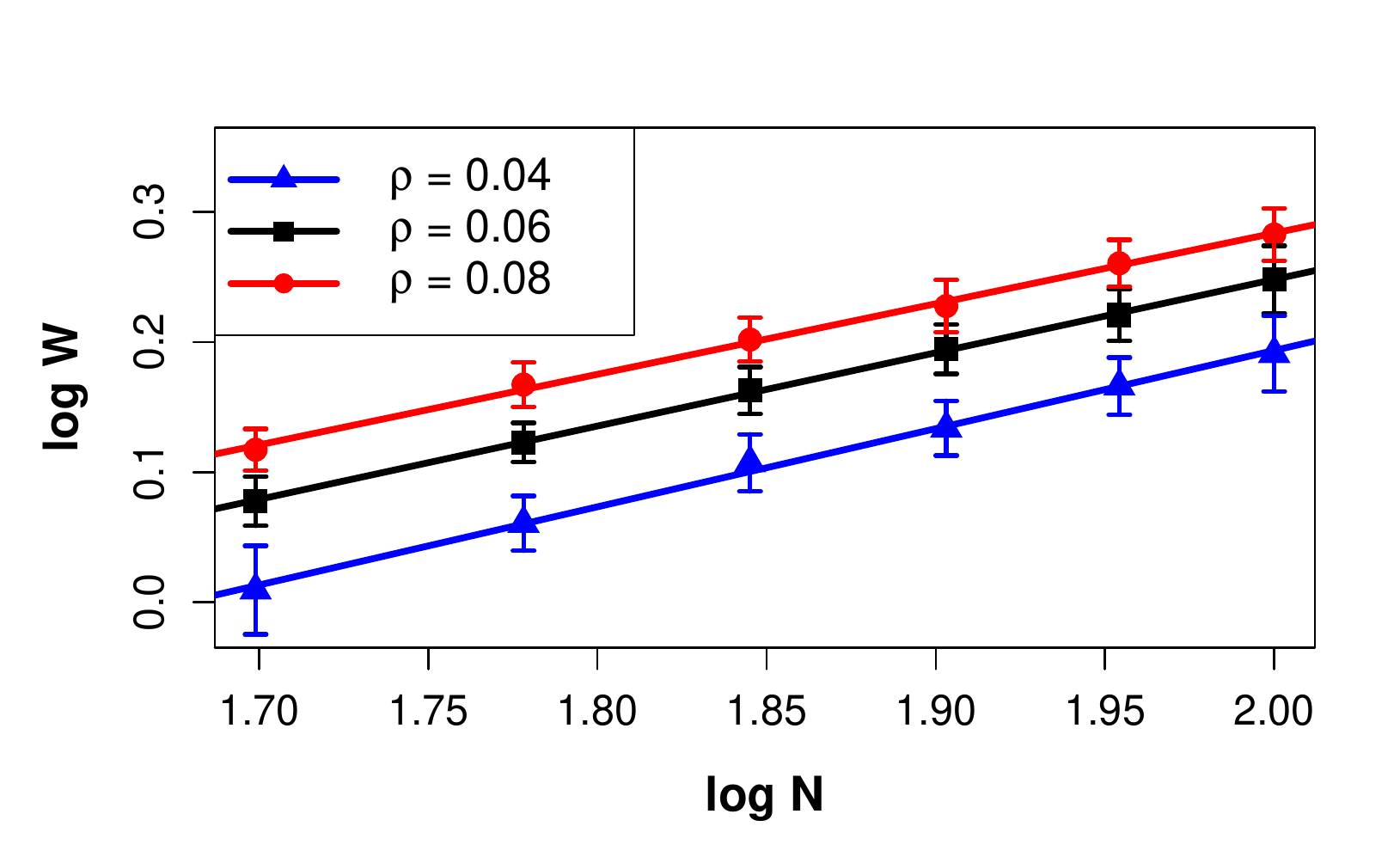}
\end{center}
\caption{Roughness of the cluster interface as a function of the
  system size $N$ for flat boundary configurations, using
    $\rho=0.04,0.06$ and $0.08$. Each value represents the average of
  50 runs over 150000 MC steps}
\label{FIG5}
\end{figure}

Even when the two clusters are well formed the interface between them
is not static. Since the boundary tends to be flat in
  average, its roughness, $W$, is defined as
\begin{equation}
  W=\sqrt{\frac{1}{N_b}\sum_i (h_i-\langle h\rangle)^2},
  \label{eq:14}
\end{equation}
where $N_b$ is the total number of border vacancies, $h$ its
  height and $\langle h \rangle$ the location of the average flat
  line. The summation index $i$ runs over all the vacancies in the
interface. This roughness typically scales with system size, $W\sim
N^\alpha$, where $\alpha$ is usually called the {\em roughness
  exponent} \cite{key-15p}, as we can see in Fig. \ref{FIG5}.

The interface is subject both to a smoothing effect and a random
noise. In other words, the system will tend to flatten the boundary in
order to minimize the dissatisfaction of the agents, yet agent
relocations are random events that will disturb the shape of the
interface. The balance between both effects is reminiscent of the
Edward-Wilkinson (EW) and Kardar-Parisi-Zhang (KPZ) universality
classes \cite{key-15p}. For the random deposition model (RD) there is no surface relaxation process.
It is interesting to estimate $\alpha$, the roughness exponent, a value that characterizes the roughness of the $saturated$ interface. For both EW and KPZ $\alpha=1/2$. All our estimates are close to this value:
$\alpha=0.600\pm0.016, 0.562\pm0.006$ and $0.542\pm0.014$ for
$\rho=0.04, 0.06$ and $0.08$, respectively (see Fig. \ref{FIG5}). 
Since RD allows the interface roughness $W$ to grow indefinitely  ($\alpha=\infty$), this class is discarded \cite{key-15p}. 

\begin{figure}[H]
\begin{center}
\includegraphics[width=8cm]{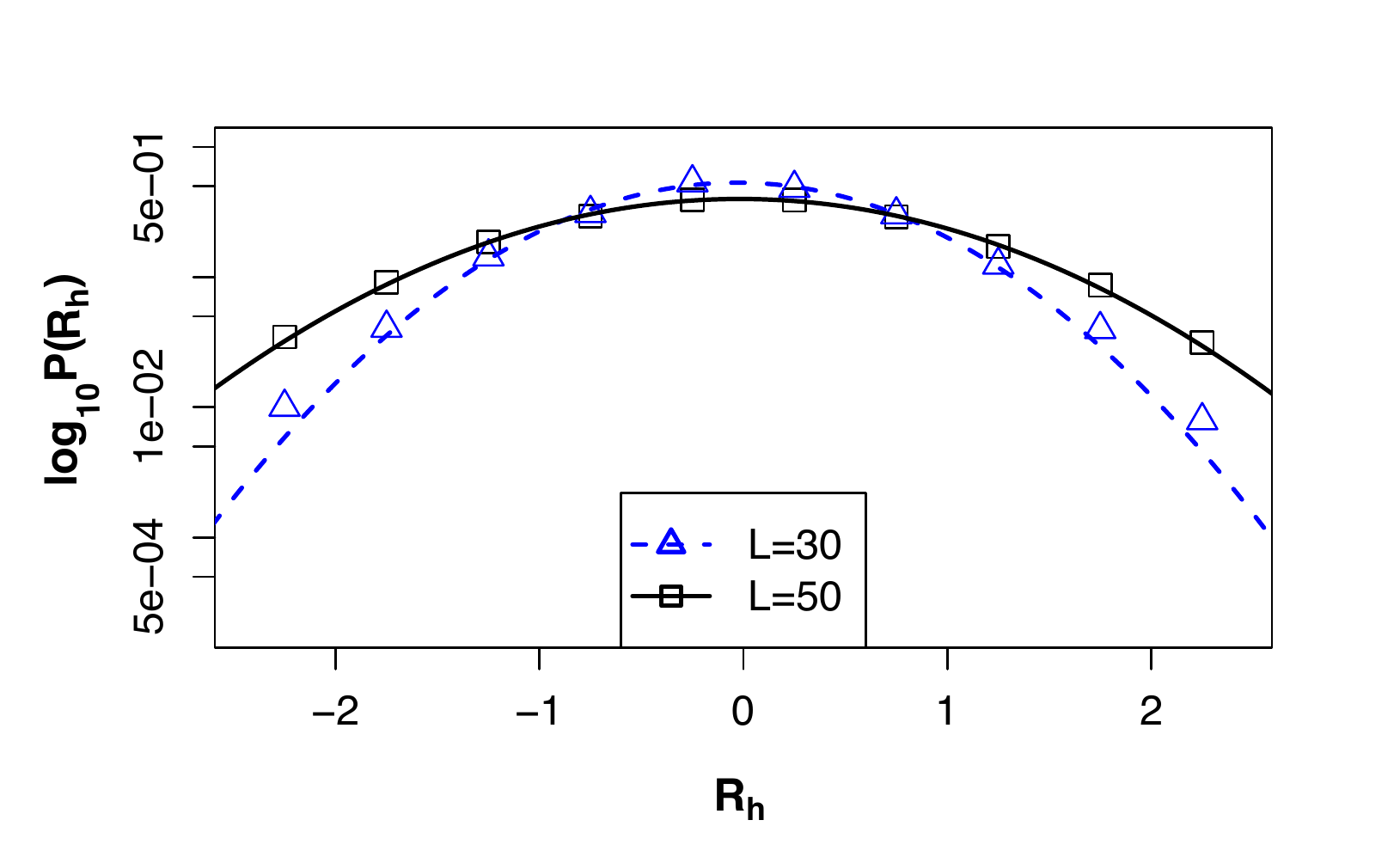}
\end{center}
\caption{Height fluctuations distributions $R_{h}$ for two window sizes: $L=50$ and $L=30$, represented as black squares and blue triangles, respectively. Both series are fitted by Gaussian distributions (solid and dashed lines). System size is $N=100$ and number of MC time steps is $1.5 \times 10^5$. }
\label{FIG6}
\end{figure}

Now we characterize the height fluctuations, $R_{h}$, for a given point $i$. These values can be calculated as:

\begin{equation}
R_{h}=h_{i} - \langle h \rangle _{L},
\label{eq:15}
\end{equation}
where $h_{i}$ is the height at point $i$ in the vacancy border and $\langle h \rangle _{L}$ is the average height of a window of total width $L$, centered around $i$. This distribution is Gaussian in the EW class \citep{edwards_wilkinson_1982} and follows a Tracy-Widom or a Baik-Rains into the KPZ one \citep{halpin-healy_kpz_2015}. As we can see from Fig. \ref{FIG6}, these fluctuactions are well fitted by Gaussians distributions for different $L$ sizes. Therefore, the border belongs to the EW universality class.

Measures of the $growth$ $exponent$ $\beta$ \cite{key-15p}, which characterizes the time-dependent dynamics of the roughening process,  are useful in the classification into an universality class. However, in our case, this parameter is difficult to estimate due to the fast creation of the border. 

\begin{figure}[H]
\begin{center}
\includegraphics[width=8cm]{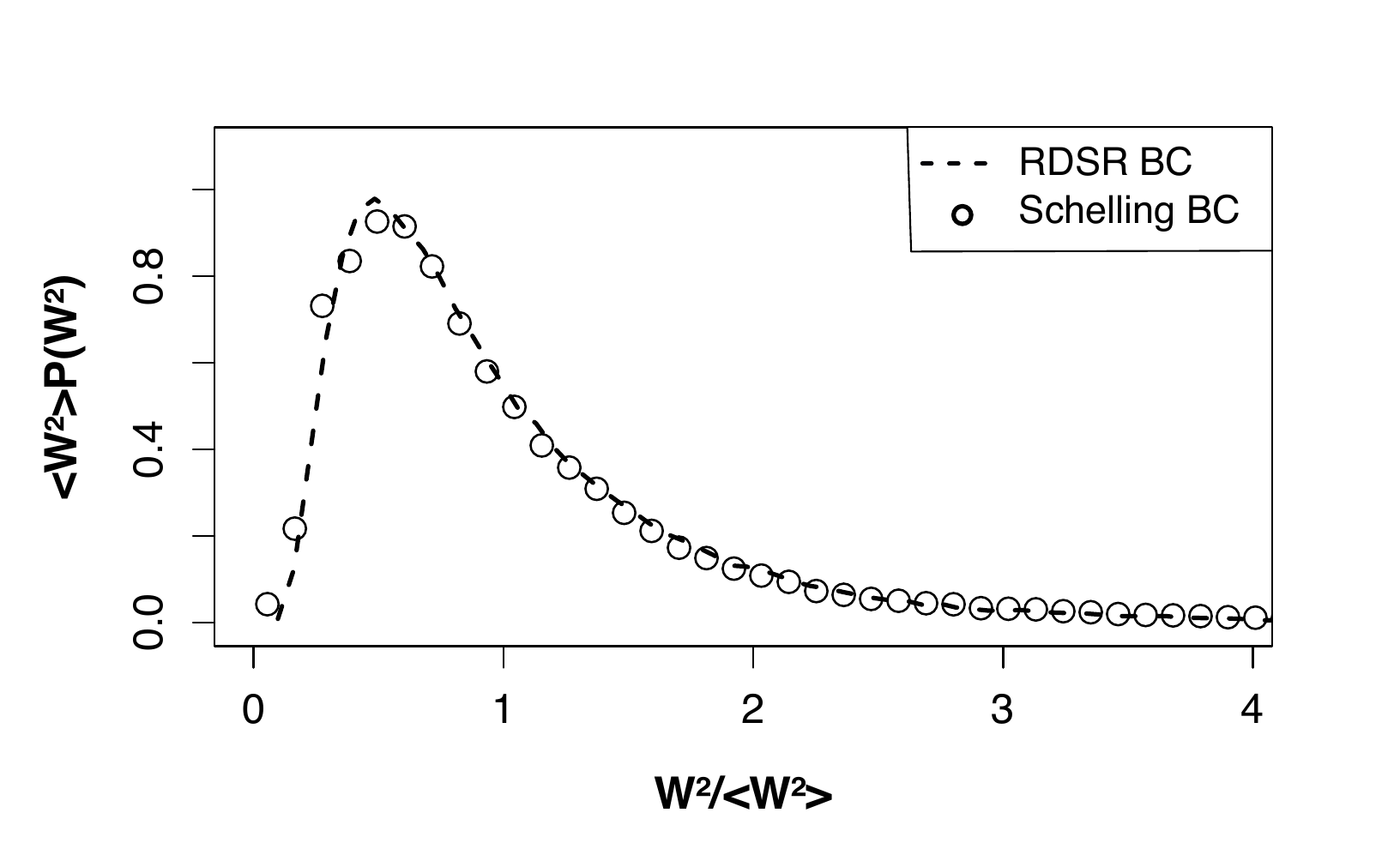}
\end{center}
\caption{Universal scaling function in its natural axis for the random surface with relaxation model and our Schelling version. System size $N=50$ and the number of MC time steps is $1.5 \times 10^5$. The boundary condition for the Schelling model is imposed in the direction perpendicular to the border at the same time step that the cooling process takes place.}
\label{FIG7}
\end{figure}

Additionally, we have compared the results of the Schelling model with those from a random deposition model with surface relaxation (RDSR) \cite{family_scaling_1986, key-15p}, a discrete model which belongs to the EW class. We have studied the distribution of the squared roughness $W^2$ \cite{queiroz_2005,rosso_2003,plischke_1994}, and estimated its probability density function of $P(W^2)$. After that we calculated the universal scaling function $\Phi(W^2/\langle W^2 \rangle)=\langle W^2 \rangle P(W^2)$ \cite{foltin_1994}, being $\langle W^2 \rangle$ the average value of $W^2$. We have solved the Schelling model and the RSDR with periodic boundary conditions and equal interface sizes. Fig. \ref{FIG7} shows the high correlation between RSDR and the Schelling model.

From a social point of view, this roughness may describe the tension between two intolerant groups in an enclosed location. Conflict will be created randomly, being dued to the system noise, whereas the agents of both groups try to minimize it flattening the border.


\section{Conclusion}
\label{sec:conclusion}

Summing up, interesting results which can be correlated to
social situations emerge when changes of the tolerance in the closed city model are  considered.

We characterize the behaviour of the system when the tolerance falls
in a continuous or a sudden way, after the system has reached
equilibrium for an intermediate or high tolerance value. For the
continuous decay we have found that the final clusterization degree of
the system, measured via the segregation coefficient, depends on the
drop rate: for slow rates the system has enough time to create
large clusters, so the segregation coefficient is close to
unity (Fig. \ref{fig1}). Thus, we may claim that along a slow
sustained decay in mutual understanding, two clusters will emerge,
minimizing the contact between them. As the process becomes faster,
a larger number of smaller cluster arises, increasing the contact
area between different groups. In this situation social frictions
can also increase.

The main feature is that once the system has reached equilibrium for
$T=1/2$ a sudden drop in tolerance $(T=1/4)$ creates a vacancy border
between the two big remaining clusters (Fig. \ref{FIG4}). The
  interface rugosity, $W$, scales with the system length $N$ as $W\sim
  N^{\alpha}$, with estimated values for $\alpha=0.600, 0.562$ and
  $0.542$ corresponding to $\rho=0.04, 0.06$ and $0.08$,
  respectively (Fig. \ref{FIG5}). The equilibrium state is not static and from a social
point of view it might describe the tension between two intolerant
groups in an enclosed location.

In addition to the estimation of the $\alpha$ exponent, the vacancy border has been characterized by several methods: height fluctuations (Fig. \ref{FIG6}) and the $P(W^2)$ distribution (Fig. \ref{FIG7}). We have found that the statistical properties of this border are close to those of the to the random deposition with surface relaxation model, and consequently, can be categorized into the Edward-Wilkinson universality class. 

Nonetheless, this approach presents some handicaps. The decision to
accept or reject a change does not take into account the agent current
happiness level, which does not seem realistic. Usually, happy people
do not want to move to another location. Besides, the characterized
system corresponds to the situation with a flat interface. Yet, a
different disposition of the clusters is possible: one of the agent
types may concentrate around a system corner developing a circular
border. The analysis of this situation goes beyond the aim of the present
paper.

Further studies should focus on variants in which agents consider
their future happiness perspectives \cite{key-25p}, or the influence
of altruistic behaviour \cite{key-26p} and the balance between
cooperative and individual dynamics \cite{key-27p}. The transfer rules
from these works combined with an open city model, where agents can leave or enter the lattice \cite{key-9p}, could lead to a framework where relocations inside and outside the city could be analyzed during several stages.

\bibliography{closed2}

\begin{appendix}

\section{Connection with the BEG model}
\label{app}
In our interpretation of the BEG model spin values can be associated
with {\em blue} agents ($s_i=+1$), {\em red} agents ($s_i=-1$) and
vacancies ($s_i=0$). Now, let us establish the connection between our segregation model and the Hamiltonian of the BEG model. 

The number of similar and different neighbors can be easily obtained from the spin variables
of sites neighboring $i$,

\begin{equation}
N_s(i)-N_d(i)=s_{i} \sum_{\<i,j\>}s_j,
\label{eq:A1}
\end{equation}

\begin{equation}
  N_s(i)+N_d(i)=s_{i}^2\sum_{\<i,j\>}s_j^2,
  \label{eq:A2}
\end{equation}
where the sum over $\<i,j\>$ should be understood as a sum over all
$j$ which are neighbors of $i$. Substituing Eq. \eqref{eq:A1} and
\eqref{eq:A2} into Eq.\eqref{eq:2}, we can rewrite our condition for
the satisfaction of agent $i$, $I_{dis} (i) \leq 0$, as

\begin{equation}
  -s_i\sum_{\<j\>}s_j-
  \(2T-1\)s_i^2\sum_{\<j\>}s_j^2\leq0,
  \label{eq:A3}
\end{equation}
where $j$ runs over his/her eight closest neighbors in the Moore
neighborhood. If an agent is allowed to move from a
site where he is not satisfied to an empty site where he
is, for a constant $T$ value, one can check that the Schelling dynamics admits as a decreasing function:

\begin{equation}
   -\sum_{\<i,j\>}s_is_j-(2T-1)\sum_{\<i,j\>}s_i^2s_j^2.
  \label{eq: A4}
\end{equation}
The Blume-Emery-Griffith model \cite{key-11p} was introduced
to study the behaviour of $He^3 -He^4$ mixtures. In this model the spin values considered are $s_i =0, \pm 1$. The
Hamiltonian can be written as:
\begin{equation}
  \mathcal{H}=-J\sum_{\<i,j\>} \,s_{i}s_{j}
  -K\sum_{\<i,j\>} \,s_{i}^{2}s_{j}^{2},
\label{eq: A5}
\end{equation}
where $\<i,j\>$ stands for the eight nearest neighbors ({\em{Moore
neighborhood}}).This Hamiltonian represents a spin-1 BEG model with
coupling constant $J$ with a biquadratic exchange constant $K$. A positive value of $J$ yields a
negative energy for each pair of neighboring agents of the
same type, while a positive value of $K$ assigns a negative
energy to every pair of neighboring agents, disregarding
their type

Comparing Eq. \eqref{eq: A4} and Eq. \eqref{eq: A5}, the former can be understood as a BEG Hamiltonian with coupling
constant $J=1$ and a biquadratic exchange of strength $K=2T-1$.

\end{appendix}

\end{document}